\documentclass[12pt]{article}
\newtheorem{defi}{Definition}

\newtheorem{prop}{Property}

 \usepackage{graphics}
\usepackage{graphicx}

\usepackage{amssymb}
\usepackage{natbib}

\begin{document}
\title{Inhomogeneous K-function for germ-grain models}


\author{M. \'Angeles  Gallego$^{(1)}$, M. Victoria Ib\'a\~nez$^{(2)}$ and Amelia Sim\'o$^{(2)}$. \\
\small{(1) Departament de Matem\`{a}tiques. Universitat Jaume I. Spain.} \\
\small{(2) Departament de Matem\`{a}tiques-IMAC. Universitat Jaume I. Spain.}}

\maketitle


\begin{abstract}
In this paper, we propose a generalization to germ-grain models of the
inhomogeneous K-function of Point Processes. We apply them to a sample of images of
peripheral blood smears obtained from patients with Sickle Cell Disease, in order to
decide whether the sample belongs to the thin, thick or morphological
region.

\textbf{keyword} Germ-grain models; K-function; Binary images; Sickle Cell Disease

\end{abstract}


\section{Introduction}
Nowadays, digital images of different phenomena of interest are commonly observed in almost every experimental fields, however, their subsequent post-processing and use tend to be very simple. The information from each image is frequently resumed in just a few numbers, and simple statistical procedures such as hypothesis testing or parameter estimation procedures are applied to them. In this paper we will work with binary images of random clumps.
Stochastic-geometric models and in particular random closed sets \cite{Matheron75,Stoyanetal95,molchanov97,molchanov05}
have been broadly used to model irregular random patterns in various fields, such as communication networks \citep{Baccellietal97}, materials science \citep{Ohseretal00} or physics \citep{Mecke98} among others. They have also been useful in modeling many medical and biological problems, especially those which require the extraction of information from microscopic images. Thus, Boolean models in particular and germ-grain models in general have been extensively used to analyze binary images of random clumps in many scientific fields \citep{Stoyanetal95,Serra82,Plaza91,Margalef74, Lyman72}.

\subsection{Germ-grain models}\label{nonhomo}

Let $\mathcal{F}$ be the class of closed subsets in the Euclidean space $I \! \! R^{d}$ and $\sigma_f = \sigma(\mathcal{F}_K, K$ compact subset of $I \! \! R^{d})$ where $\mathcal{F}_K =\{F \in \mathcal{F}: F \cap K \neq \varnothing \}$ or equivalently the Borel $\sigma$-algebra generated by the
 the Fell topology on $\mathcal{F}$ \citep{Fell62}. If $\textit{P}$ denotes a probability measure in $(\mathcal{F}, \sigma_f)$ then, according with the original definition given by  \cite{Matheron75}, $(\mathcal{F}, \sigma_f, \textit{P})$ is a random closed set.

The germ-grain model \citep{Hanisch81} is a particular class of random closed set model that has proved to be suitable for working with images formed by random clumps. Its mathematical definition is as follows:

\begin{defi}\label{Def germ-grain model}
Let $\Psi=\{[x_{n};\Xi_{n}]\}$ be a marked point process, where $x_{n}$ are points of $I \! \! R^{d}$ and  $\Xi_{n}$ are compact subsets of $I \! \! R^{d}$. A germ-grain model, $\Xi$, is defined as:
\begin{displaymath}
    \Xi =\bigcup_{n}(\Xi_{n}+x_{n}).
\end{displaymath}
The points $x_{n}$ are called \emph{germs} and the sets $\Xi_n$ are known as \emph{grains}\citep{Hanisch81}.
\end{defi}

The most widely known and used germ-grain model is the Boolean model \citep{molchanov97}. The Boolean model is obtained when the germ process, $\phi_\lambda=\{x_1,x_2,\cdots \}$, is a Poisson process with intensity function $\lambda(x)$ and the grains, $\Xi_{0}$, $\Xi_{1}$, $\Xi_{2}$, $\cdots$, are i.i.d. and independent of the germs.

Under homogeneity, the intensity of the Poisson germs process is assumed to be constant, and it is denoted by $\lambda$.  Additionally, if $\Xi$ is stationary, i.e. if $\Xi$ and the translated sets $\Xi_x=\Xi+x$  have the same distribution $\forall x \in I \! \! R^{d}$, the \emph{typical grain}, $\Xi_{0}$, can be defined as a closed random set with the same distribution as the sets $\{\Xi_{n}\}$  but independent, both of them and of the germs. So, the distribution of the \textit{typical} grain is the distribution of the marks of the marked point process $\Psi=\{(x_{n}, \Xi_{n}): n \in \mathbb{N}\}$. This is a marked point process on $ I \! \! R^{d}$, with the marks $\Xi_{n}$, $n \in \mathbb{N}$, being random convex bodies in $ I \! \! R^{d}$. It is a distribution on the space $\mathcal{K}$ of compacts sets.

The probability distribution of a random closed set $\Xi$ is given by its \textit{capacity functional}, $T_{\Xi}$, defined as:
  \begin{eqnarray*}
  T_{\Xi}(K)=P(\Xi \cap K \neq \emptyset) \mbox{\hspace*{0.5cm}   } \forall K \subset I \! \! R^{d}.
  \end{eqnarray*}

Useful summary descriptions of the probability distribution of the random closed set are the coverage function, $p(x)$, defined as the mean area of $\Xi$
in the unit square centered in $\{x\}$, i.e. $p(x)=T_{\Xi}(\{x\})$, and the covariance function $C(x,x+h)=T_{\Xi}(\{x, x+h\})$. These functions describe the first-order and the second-order structure of the set respectively.

Under stationarity the coverage function is constant, $p(x)=p(0)=p$, and is called volume fraction and $ C(x, x+h)=C(h)$ $\forall x \in I \! \! R^{d}$. In this case, the K-function \citep{Daley88, Jensenetal90} provides a more intuitive and practical way to describe the second order structure. It is defined as:
\begin{displaymath}
   K(t)=\frac{1}{p}E_0[\nu(\Xi \cap B(0,t)],
\end{displaymath}
where $B(0, t)$ is the ball centered at the origin with radius $t$; $\nu$ is the Lebesgue measure and $E_0$ denotes the expectation with respect to the Palm distribution of $\Xi$. When $\{0\}$, the origin of $I \! \!R^{d}$ is chosen as the typical point of  $\Xi$, $pK(t)$ becomes the average measure of the intersection of $\Xi$ with a ball of radius $t$ centered at  $\{0\}$.

We have the following relation between the covariance and the $K$-function:
\begin{displaymath}
   K(t)=\frac{1}{p^2}\int_{B(0,t)}C(h)dh.
\end{displaymath}

The $K$-function, also called the Ripley $K$-function, has been extensively used in the point processes literature \citep{Ripley77, Diggle83} mainly to analyze the strength of the interaction between points in the point process. However, it has also been used to analyze isotropic Boolean models \citep{ayalaetal95, ayalaetal98}. Additionally, \citet{ayalaetal93} proposed an approximation for the K-function in overlapping Boolean models based on an approximation of the covariogram of the primary grain.
 Originally, the K-function was defined to characterize stationary point processes, but the definition was later extended to inhomogeneous point processes \citep{Baddeleyetal00,Diggleetal2007}. As far as we know, it has not yet been extended to non-stationary germ-grain models. This is our objective in this paper.

The assumptions of stationarity and isotropy facilitate the estimation of the parameters of the germ-grain model. However, the hypothesis of spatial homogeneity frequently fails when real data sets are analysed. An importante example of non stationary germ-grain model is the non-homogeneous Boolean model, i.e. the Boolean model  obtained when the germ process is a Poisson process with intensity function $\lambda(x)$. Non-homogeneous Boolean models have been used to model functionally graded materials \citep{Hahnetal99,QuintanillaTorquato97}, distributions of galaxies \citep{BondLevPogosyan95} and complex fluids \citep{Brodatzki01}. Methods to estimate parameters of  non-homogeneous Boolean models have been studied by \cite{molchanov00} and by \cite{Schmitt97}.

From now on we will restrict our work to the case $d = 2$.

To define the  inhomogeneous K-function for point patterns, \citet{Baddeleyetal00} considered a point process $Y$  in  $I \! \! R^2$, with first-order intensity function $ \lambda(s), s \in I \! \! R^2$. Given $\mathfrak{B_0}$ the class of bounded Borel sets in $I \! \! R^2$, and assuming that the function
\begin{eqnarray}
M(A,B)=E\sum_{y_i \in Y\bigcap A}\sum_{y_j \in Y\bigcap B}\frac{1}{\lambda(y_i)}\frac{1}{\lambda(y_j)}
\end{eqnarray}
is finite for all $A,B\in \mathfrak{B_0}$, they defined $Y$ as a second-order intensity reweighted stationary point process if $M(A,B)=M(A+x,B+x)$, being $A+x$ the translation of $A$ by the vector x. From this point, they defined the inhomogeneous K-function for point processes as:

\begin{defi} [Inhomogeneous K-function for point processes] \label{Def inhomogeneous K-function}
Let $Y$ be a second-order intensity reweighted stationary point process. Then, the inhomogeneous K-function of $Y$ is defined as:
\begin{equation}
 \label{Kin} K_{inhom}(t)=\frac{1}{\nu(W)}\mathbf{E}\sum_{y_{i}\in Y\cap W} \sum_{y_{j}\in Y \backslash\{y_{i}\}}\frac{1(\|y_{i}-y_{j}\|\leq t)}{\lambda(y_{i})\lambda(y_{j})}, \ t\geq 0; \\
\end{equation}
for any $W \in \mathbf{B}_{0}$, the class of bounded Borel sets in $I \! \! R^2$,  where $1( \cdot)$ denotes the indicator function, $\nu(W)$ is the area (Lebesgue measure) of $W$, , and  $a/0:=0$ for $a \geq 0$. This expression does not depend on the choice of $W$.
\end{defi}

Given a realization of $Y$ in an observation window $W$, its corresponding sample estimator \citep{Baddeleyetal00} becomes:

\begin{equation}
 \label{Kin2} \hat{K}_{inhom}(t)=\frac{1}{\nu(W)}\sum_{y_{i}\in Y\cap W}  \sum_{y_{j}\in Y \backslash\{y_{i}\}} \frac{w_{y_i,y_j} 1(\|y_{i}-y_{j}\|\leq t)}{\hat{\lambda}(y_{i})\hat{\lambda}(y_{j})}, 0\leq t \leq t^*, \\
\end{equation}
where $w_{y_i,y_j}$ is an edge corrector function and $t^*=sup\{r \geq 0: \nu(\{s \in W: \partial B(s,r) \cap W \neq \emptyset \})>0\}$, where $\partial B(s,r)$ denotes the boundary of $B(s,r)$.

\section{K-function for inhomogeneous germ-grain models}
As stated below, the inhomogeneous K-function has been defined to work with inhomogeneous point processes \citep{Baddeleyetal00,Diggleetal2007}, and our aim is to extended it for non-stationary germ-grain models. We have to note that there is an important difference between point processes and germ grain models. The probability distribution of a point process is characterized by its random count measure, but the random coverage measure of a random closed set does not characterize its probability distribution. (\cite{AyalaFerrandizMontes91}), establish a sufficient condition to guarantee that the random coverage measure determine the probability distribution of the random closed set.

Previously to define the inhomogeneous K-function, the following fundamental concept must be introduced.

Assuming that $\Xi$ has a strictly positive coverage function, i.e. $p(x)>0$ $\forall x$, given $A$, $B$ bounded Borel sets in $\mathbb{R}^2$, the measure $M$ is defined in $\mathbb{R}^4$ as:
\begin{displaymath}
  M(A,B)=E \int_{\Xi \cap A}  \int_{\Xi \cap B} \frac{1}{p(x) p(y)}dx dy.
\end{displaymath}

\begin{defi}[Second-order intensity-reweighted stationary]
The germ-grain model $\Xi$ is "second-order intensity-reweighted stationary" if $M(A,B)=M(A+x, B+x)$ for all $x \in \mathbb{R}^2$
\end{defi}

A second-order stationary germ-grain model is also second-order intensity-reweighted stationary.
A non-homogeneous Boolean model is second-order intensity-reweighted stationary because of the second-order intensity-reweighted stationary property of its germ process and the independence of the grains.

\begin{defi} [Inhomogeneous K-function  for germ-grain models] \label{Def inhomogeneous K-function for boolean models}
Let $\Xi$ be a second-order intensity-reweighted stationary germ-grain model in an observation window $W$.  The inhomogeneous K-function of $\Xi$ is defined as:
\begin{equation}\label{Kinteorica}
K_{inhom}(t) =\frac{1}{\nu(B)}E(\int_{\Xi \cap B}\int_{\Xi \cap B(y,t)}  \frac{1}{p(x)p(y)} dy dx)
\end{equation}
for any  $B \in \sigma_f$.
\end{defi}

\begin{prop} \label{No_depende_B}
Definition \ref{Kinteorica}  does not depend on $B$.
\end{prop}

\textbf{Dem.}

Let $A_t=\{(x,y): x \in B, \ y \in B(x,t)\}$,
\begin{eqnarray*}
M(A_t)=E(\int_{\Xi \cap B}\int_{\Xi \cap B(y,t)}  \frac{1}{p(x)p(y)}) dy dx.
\end{eqnarray*}

Because of the second-order intensity-reweighted stationary property,
$M(A_t)=M(A_t+(z,z))\hspace*{0.5cm} \forall z \in I \! \!R^{2}$,
and as a result

\begin{eqnarray*}
\frac{1}{\nu(B)}E(\int_{\Xi \cap B}\int_{\Xi \cap B(y,t)}  \frac{1}{p(x)p(y)} dy dx)=\frac{1}{\nu(B+z)}E(\int_{\Xi \cap B+z}\int_{\Xi \cap B(y,t)}  \frac{1}{p(x)p(y)} dy dx)
\end{eqnarray*}

 c.q.d.

\vspace{0.4cm}

$K_{inhom}(t)$ has an interpretation as a Palm expectation, similar to that for the stationary case:

\begin{prop}
\begin{displaymath}
K_{inhom}(t)=E_s (\int_{\Xi \cap B(s,t)}  \frac{1}{p(x)} dx) \ \ \forall s \in I \! \!R^{2}.
\end{displaymath}

where $E_s $ denotes the expectation with respect to the Palm distribution $P_s$ of $\Xi$ at $s$ that can be interpreted as the conditional distribution of $\Xi$ given that $s \in \Xi$.
\end{prop}

\textbf{Dem.}

Consider $B \in \sigma_f$.

\begin{eqnarray*}
&& \frac{1}{\nu(B)} \int_{B} E_s ( \int_{\Xi \cap B(s,t)}  \frac{1}{p(x)} dx) ds= \frac{1}{\nu(B)}\int_{B} \frac{p(s)}{p(s)}E_s (\int_{\Xi \cap B(s,t)}  \frac{1}{p(x)} dx) ds =\\
&& \frac{1}{\nu(B)} \int_{I \! \!R^{2}} p(s)E_s ( 1_B(s) \int_{\Xi \cap B(s,t)} \frac{1}{p(s)p(x)} dx) ds.
\end{eqnarray*}

Applying the Campbell-Mecke formula \citep{Jensenetal90}:

\begin{eqnarray*}
\frac{1}{\nu(B)}\int_{I \! \!R^{2}} p(s)E_s ( 1_B(s) \int_{\Xi \cap B(s,t)}  \frac{1}{p(x)} dx) ds\frac{1}{p(s)p(x)} dx) ds=\frac{1}{\nu(B)}E(\int_{\Xi \cap B}\int_{\Xi \cap B(s,t)}  \frac{1}{p(s)p(x)} dx ds
\end{eqnarray*}
From the result of the Prop. \ref{No_depende_B}, we know that the second term of the equality does not depend on $B$. Then:

\begin{eqnarray*}
\frac{1}{\nu(B)}\int_{B} E_s (\int_{\Xi \cap B(s,t)}  \frac{1}{p(x)} dx) ds=\frac{\nu(B)}{\nu(B)}E_s (\int_{\Xi \cap B(s,t)}  \frac{1}{p(x)} dx).
\end{eqnarray*}

And so:
\begin{eqnarray*}
E_s (\int_{\Xi \cap B(s,t)}  \frac{1}{p(x)} dx)= \frac{1}{\nu(B)}E(\int_{\Xi \cap B}\int_{\Xi \cap B(s,t)}  \frac{1}{p(s)p(x)} dx ds)
\end{eqnarray*}

c.q.d.

\section{Estimation}

Given a realization of $\Xi$ in an observation window $W$ its corresponding sample estimator is

\begin{eqnarray}\label{Kin3}
\hat{K}_{inhom}(t)&& =\frac{1}{\nu(W)}\int_{\Xi \cap W}\int_{\Xi \cap W}  \frac{w_{y_i,y_j} 1(\|y_{i}-y_{j}\|\leq t)}{\hat{p}(y_{i})\hat{p}(y_{j})} dy_i dy_j \cong \\
&& \cong \frac{1}{\nu(W)}\sum_{y_{i}\in \Xi \cap W} \sum_{y_{j} \in \Xi \cap W}\frac{w_{y_i,y_j} 1(\|y_{i}-y_{j}\|\leq t)}{\hat{p}(y_{i})\hat{p}(y_{j})}, \ 0\leq t \leq t^*,\nonumber
\end{eqnarray}
with $w_{y_i,y_j}$ an edge corrector function, and $t^*=sup\{r \geq 0: \mid \{s \in W: \partial B(s,r) \cap W \neq \emptyset \}\mid>0\}$, as above,
 $\hat{p}$ is a kernel estimator of the coverage function, and $\cong$ denotes the numerical approximation.

\begin{figure}
   \begin{center}
\begin{tabular}{cc}
\includegraphics [width=5cm]{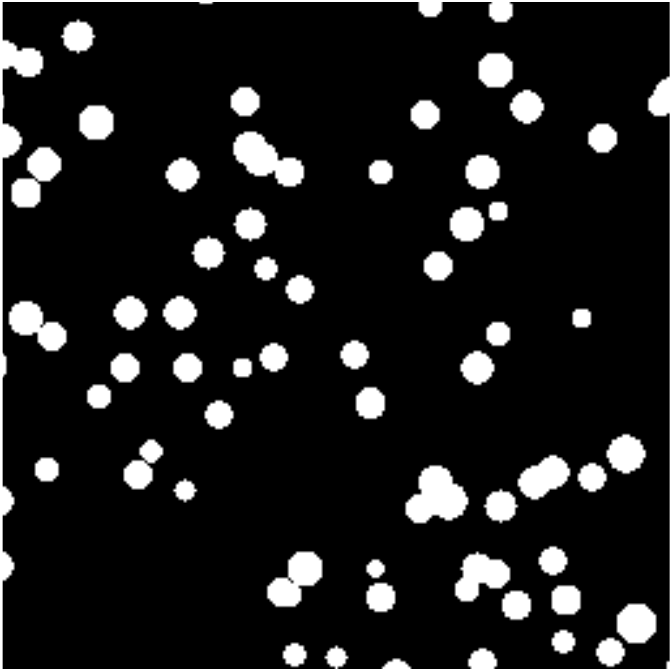} &
\includegraphics [width=5cm]{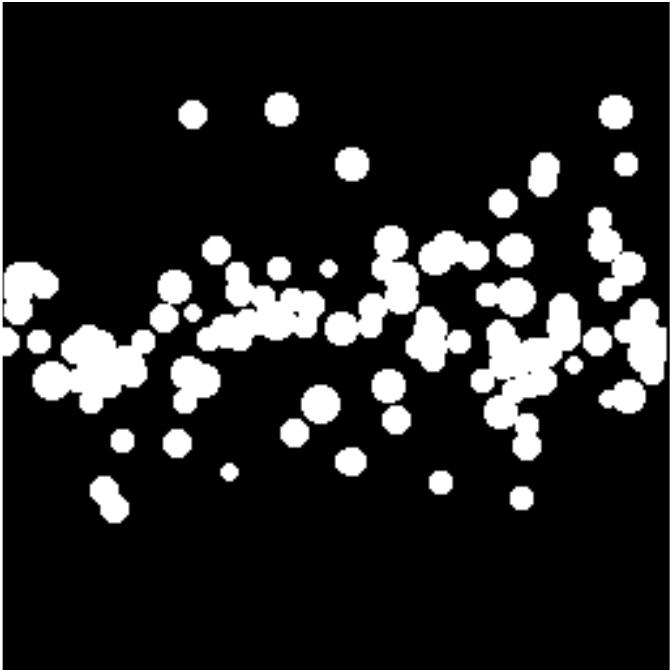} \\
(a) & (b) \\
\includegraphics [width=5cm]{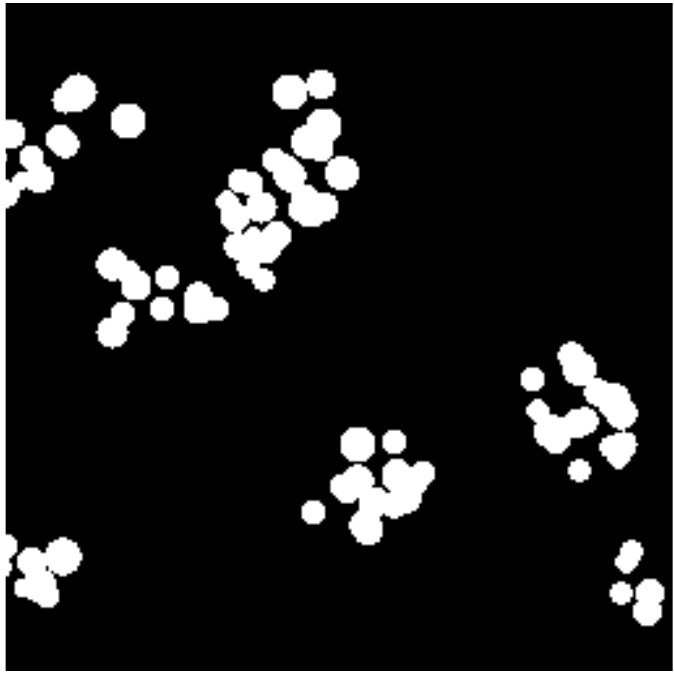} & \includegraphics [width=6 cm]{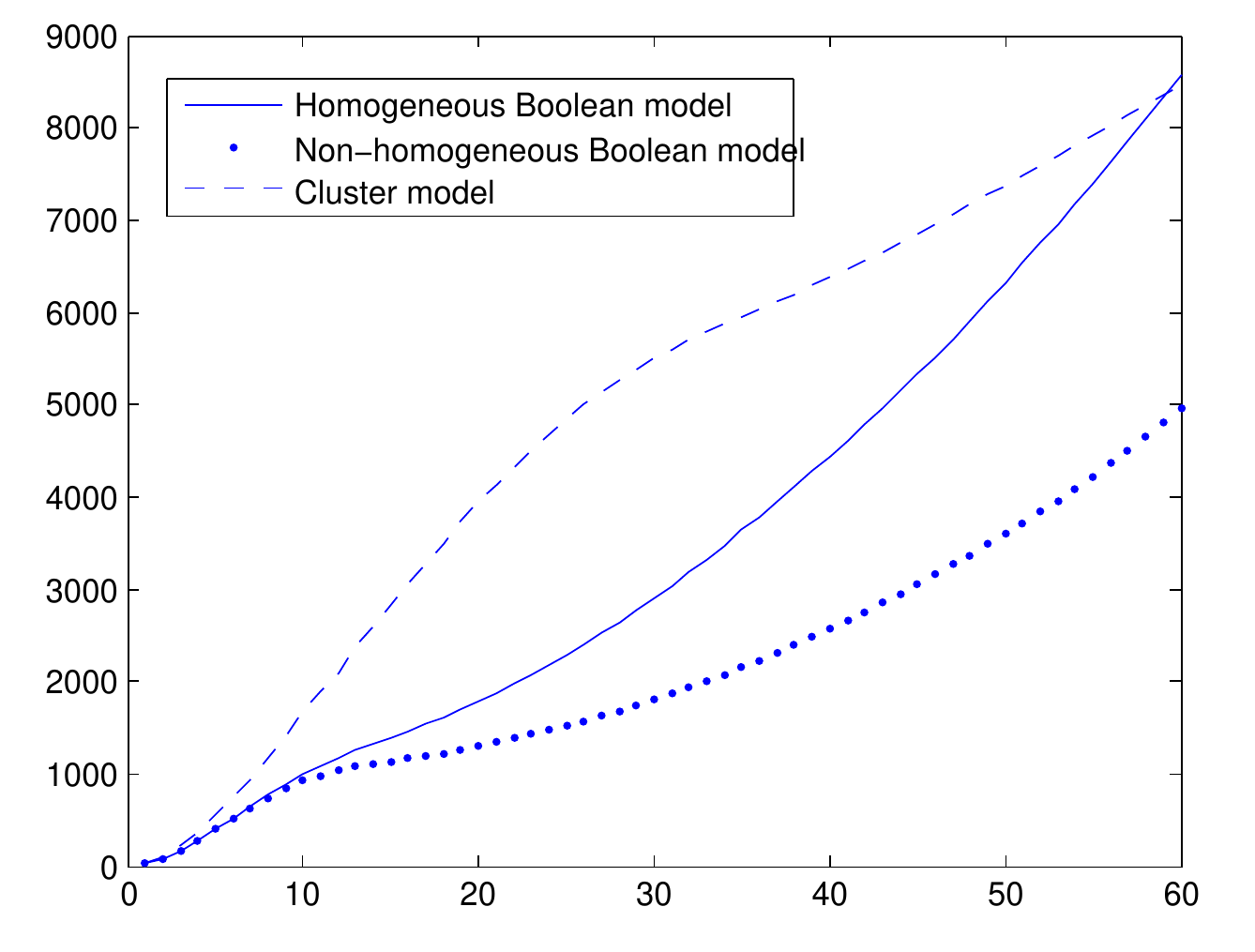}\\
 (c) & (d)\\
 \end{tabular}
  \end{center}
  \caption{ Realization of different germ-grain models: (a) Homogeneous Boolean model, (b) non-homogeneous Boolean model, (c) Cluster model, and (d) their Sample Inhomogeneous K-function \label{ejemploKt}}
\end{figure}

Figure \ref{ejemploKt} shows realizations of three different germ-grain models and the mean of the estimated inhomogeneous K-function computed from a sample of them.  It can be seen that for $t$ $\approx <60$, $\hat{K}_{inhom}(t)$ is greater for the cluster model than for the Boolean model, which is due to the effect of the clustering of germs. For  $t$ $\approx <30$  it also increases much faster but for larger $t$-values the increase in $\hat{K}_{inhom}(t)$ for the Cluster model is slower than for the Boolean model. Finally, for $t$ $\approx > 60$, $\hat{K}_{inhom}(t)$ is lower for the cluster model than for the Boolean model.

As \cite{Baddeleyetal00} state, in practice it is difficult to make a distinction between large-scale variation given by $p(x)$ and variation due to interactions. In this case it is of particular importance the choice of the bandwidth parameter in the kernel estimator of the coverage function (Eq. \ref{Kin3}). In our experiments, we choose it comparing the value of the analytical expression of the volume fraction of an homogeneous Boolean model, with the empirical one for different values of $h$. On the other hand, taking into account that the theoretical volume fraction in the Cluster model is constant we choose the greatest $h$ value among those that provided a good approximation. This choice of $h$ allows to distinguish clearly the $\hat{K}_{inhom}(t)$ for the non-homogeneous Boolean model and the Cluster model, as can be seen in figure \ref{ejemploKt} (d). It is due to the fact that the estimation of the coverage function in the "accumulation" area of the non-homogeneous Boolean model is quite greater than the  corresponding to the Cluster model.


\begin{figure}
   \begin{center}

\includegraphics [width=5cm]{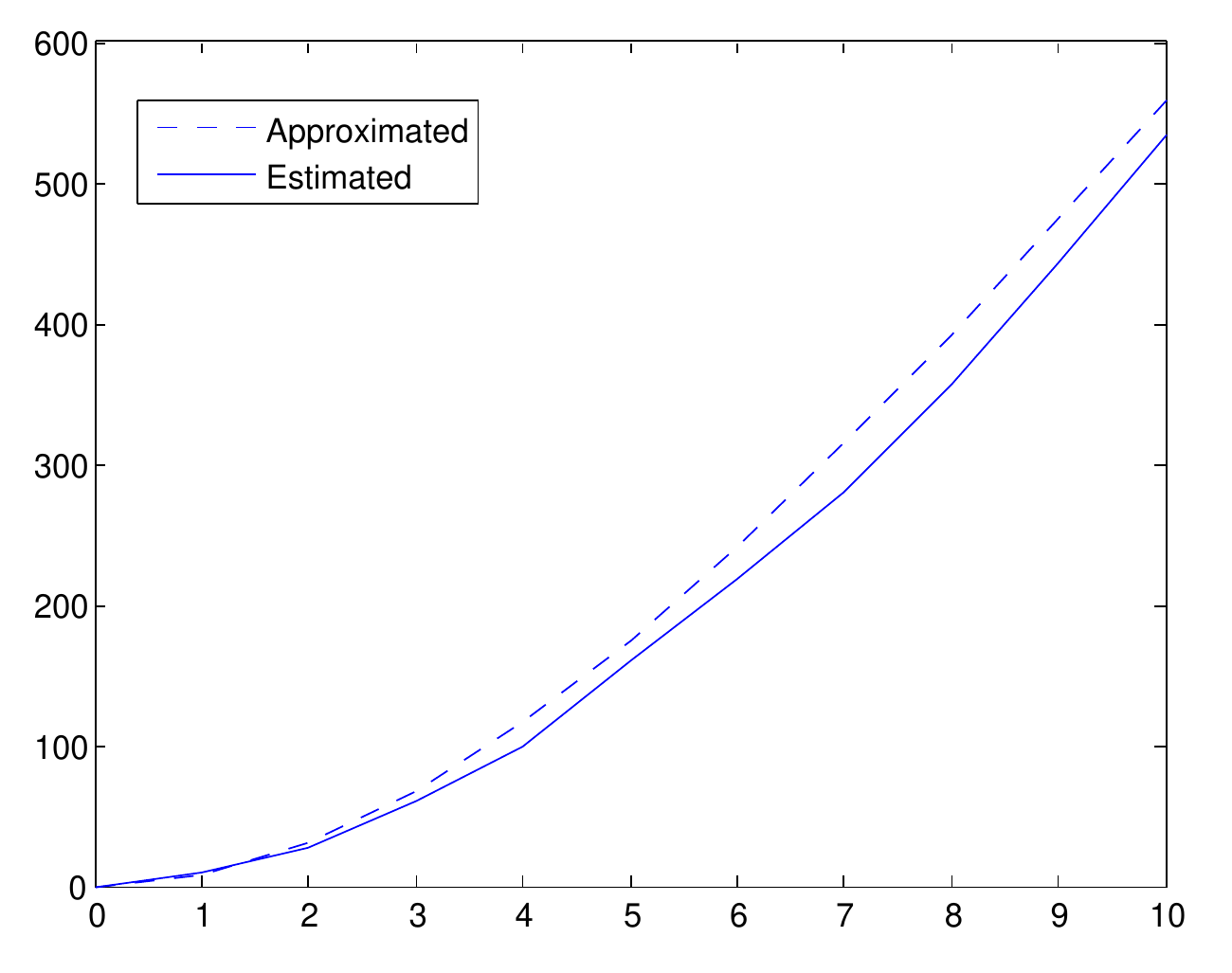}
  \end{center}
  \caption{ Mean of the estimated inhomogeneous K-function corresponding to $10$ realizations of a Boolean model and the approach given by \cite{ayalaetal93}. \label{comparacionK}}
\end{figure}

In contrast to what happens in point processes context it does not exist an exact expression for the K-function for an homogeneous Boolean model. As said above, \cite{ayalaetal93} gave an approximate expression for it based on an approximation of the covariogram of the primary grain. This approach is valid for values of $t$ close to zero.  In Fig. \ref{comparacionK}
we compare the mean of the estimated inhomogeneous K-function corresponding to $10$ realizations of a Boolean model with the approach given by \cite{ayalaetal93}.

\section{Digital images of peripheral blood smears}\label{aplicacio}

As an example of application we use the inhomogeneous K-function to perform unsupervised classification when the sample information are digital images of peripheral blood smears.

Examination of peripheral blood smears is an essential component in the evaluation of all patients with hematologic disorders.  In particular it is used in the diagnosis and monitoring of an important genetic disease called Sickle Cell Disease (SCD). SCD causes the hardening or polymerization of the hemoglobin that contains the erythrocytes.  The cells are deformed and tend to block blood flow in the blood vessels of the limbs and organs. Blocked blood flow can cause pain and organ damage. This disease has been recognized as a major public health problem by international agencies such as the World Health Organization (WHO) and the United Nations Educational, Scientific and Cultural Organization (UNESCO).
Depending on the state of the disease, i.e., depending on the quantity of deformed cells in their blood flow,  patients are classified into three groups: those with a benignant form, without pain crises, those with a moderate form who  only have one crisis per year, and those who are seriously ill with two or more crises per year.
The quantitative analysis of digital images of peripheral blood smears offers useful results in the clinical diagnosis of this illness, and guides the specialist in allocating the most suitable treatment.

\begin{figure}[h!]
   \begin{center}
\begin{tabular}{ccc}
\includegraphics [width=3.5cm]{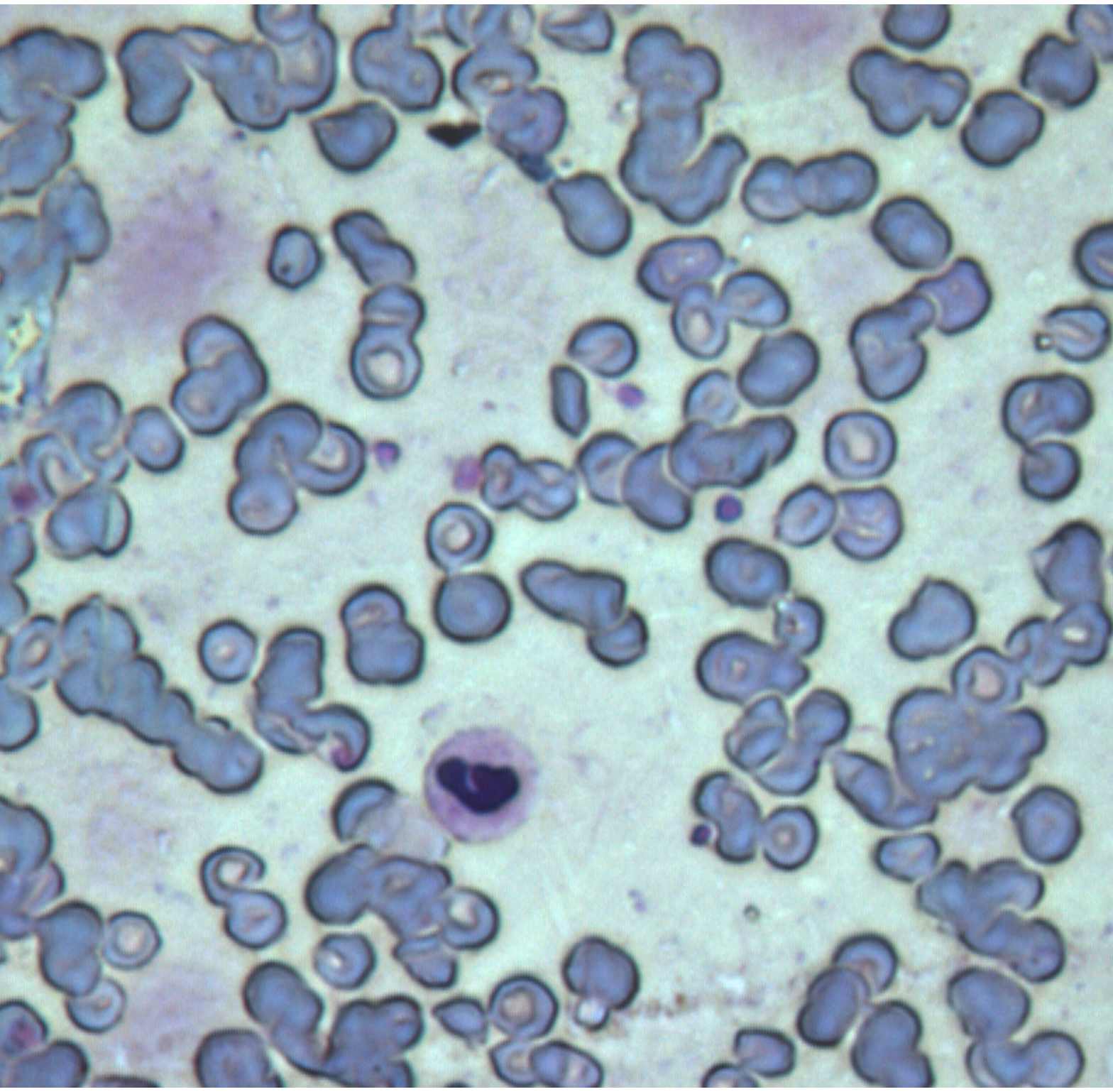} &
\includegraphics [width=3.5cm]{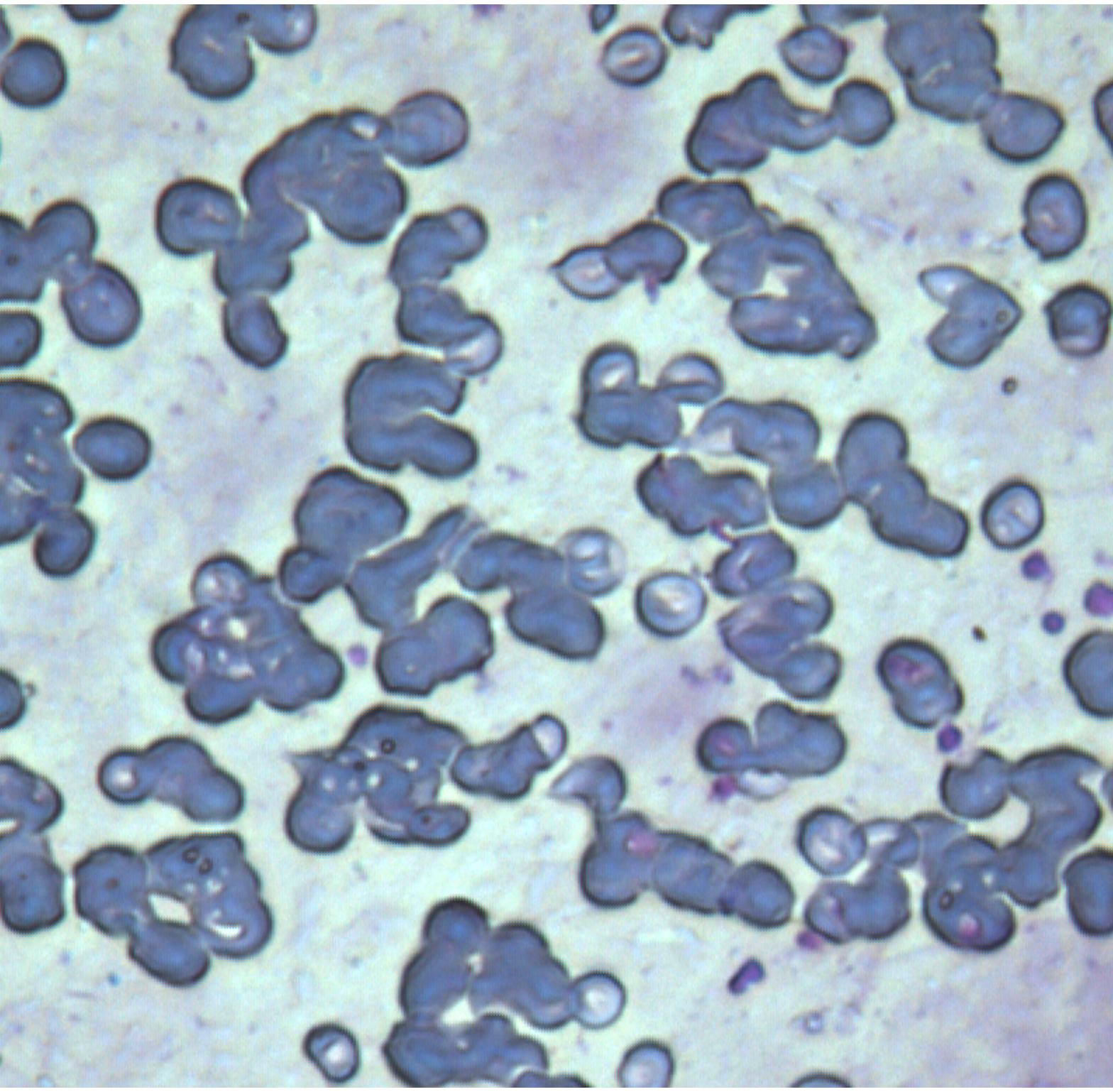} &
\includegraphics [width=3.5cm]{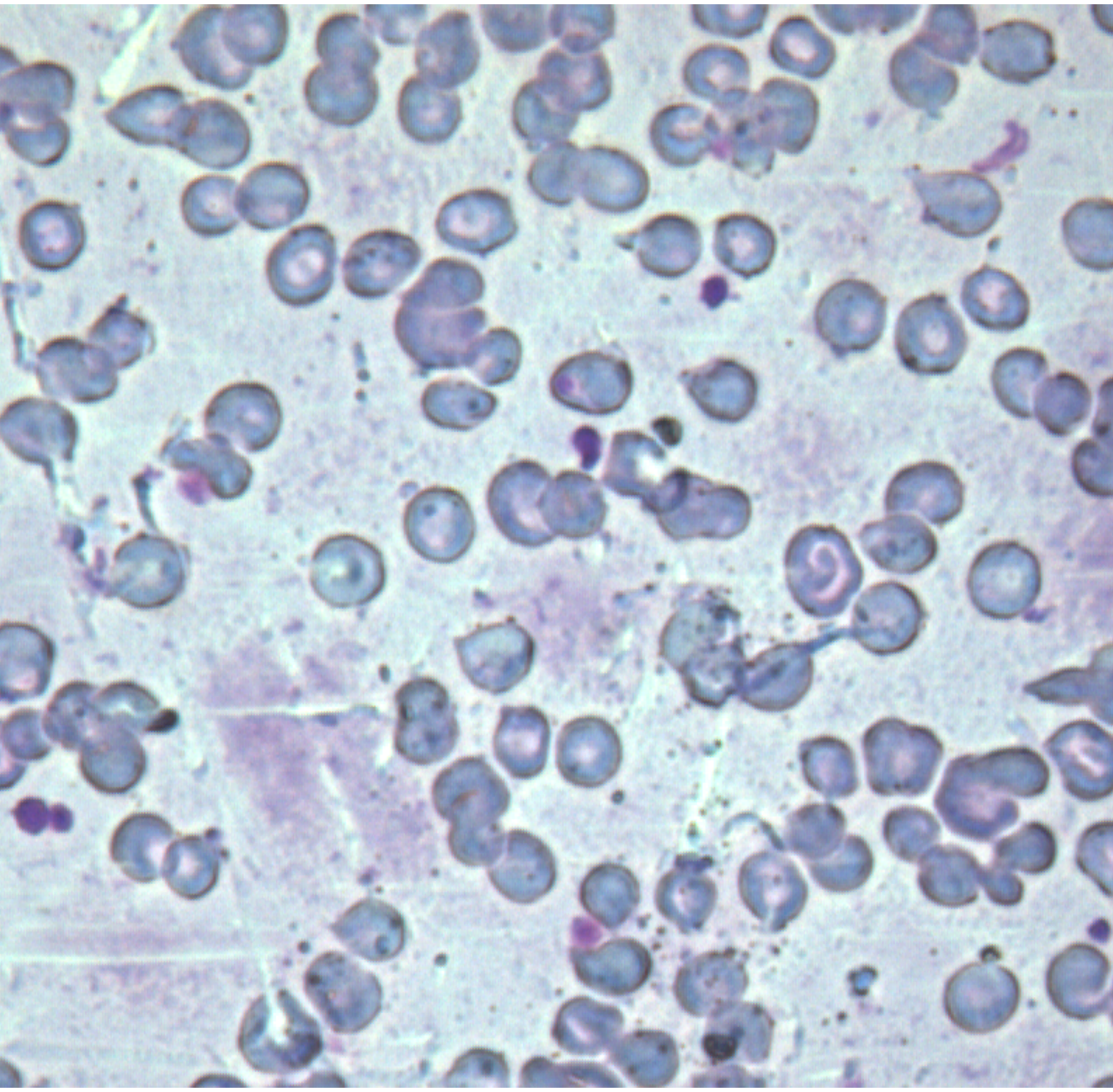} \\
\includegraphics [width=3.5cm]{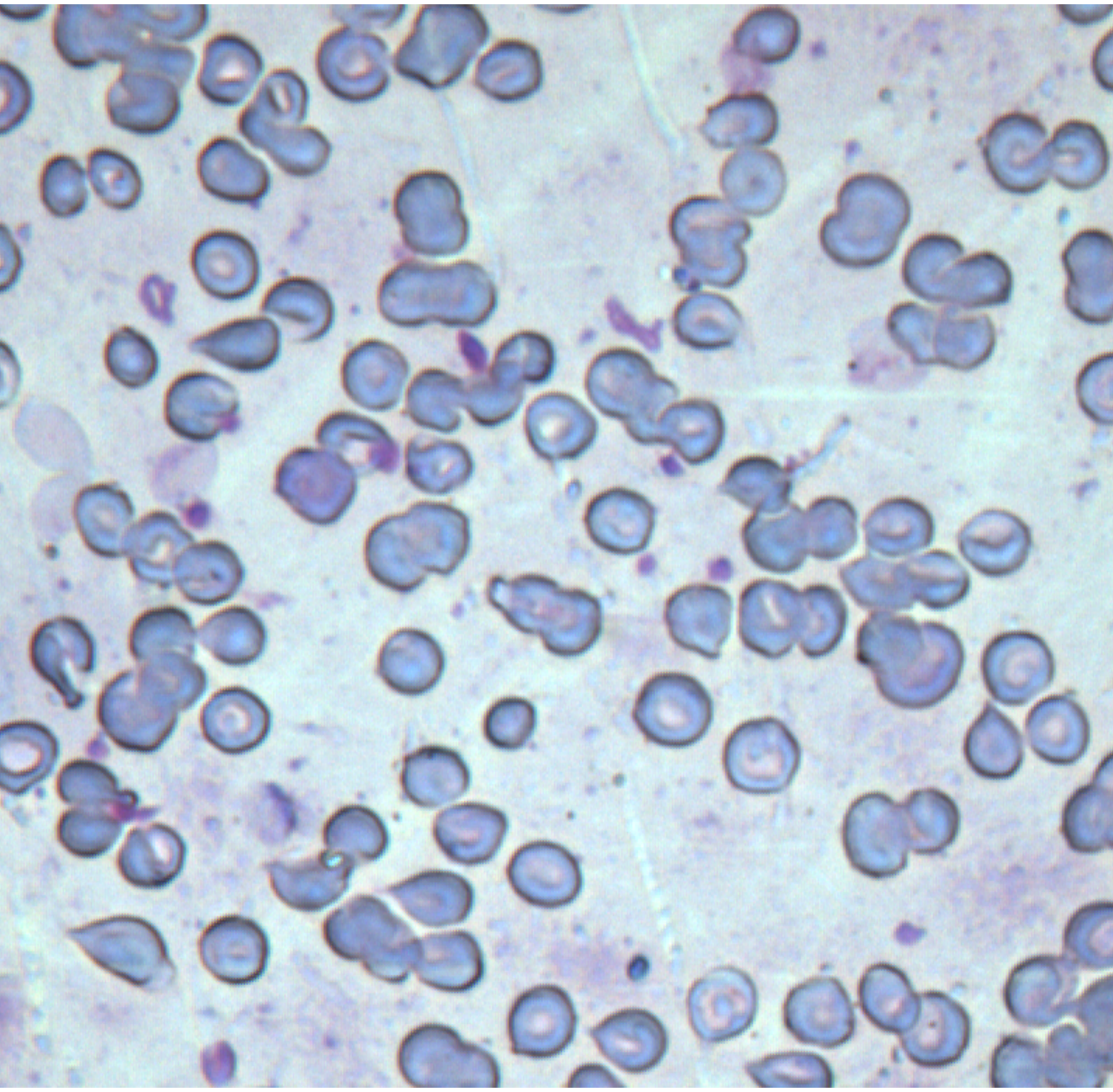} &
\includegraphics [width=3.5cm]{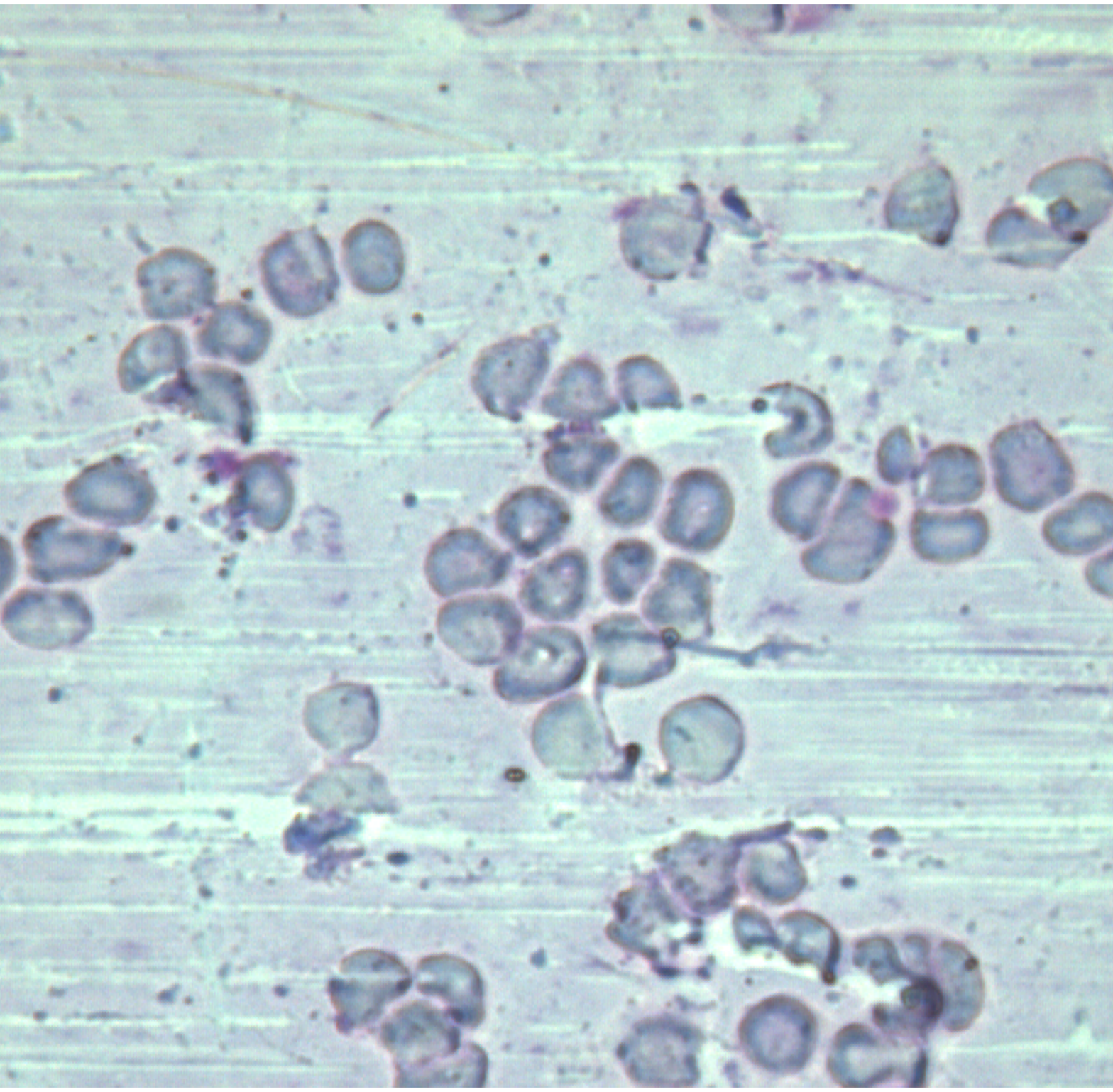} &
\includegraphics [width=3.5cm]{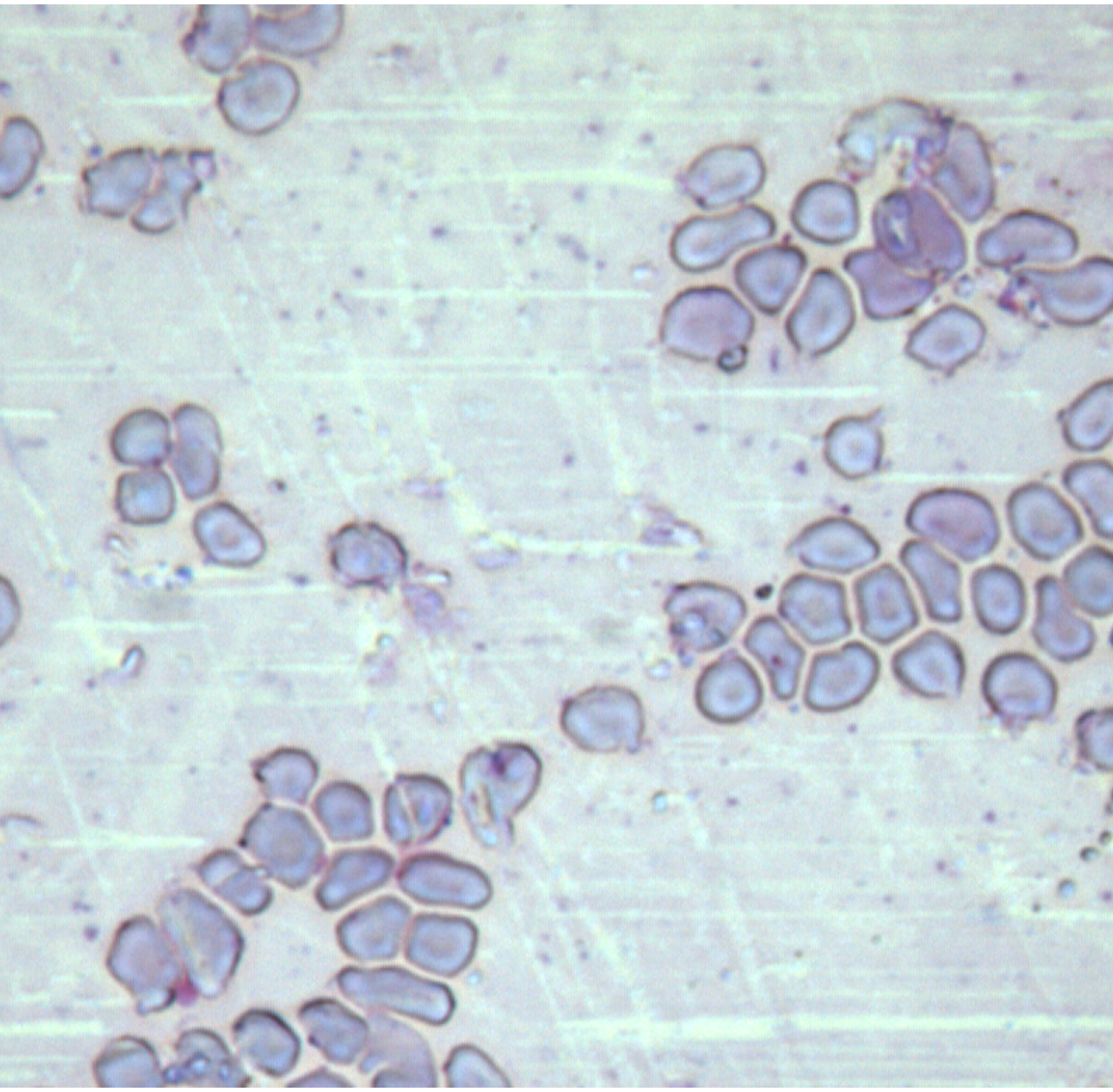} \\
 \end{tabular}
  \end{center}
  \caption{ Different images of different zones of the same peripheral blood smear \label{sangre1}}
\end{figure}

A peripheral blood smear (peripheral blood film) is a glass microscope slide coated on one side with a thin layer of venous blood. The slide is stained with a dye and examined under a microscope. A blood-smear preparation requires dropping the blood sample,
spreading the sample, and staining. Sample spreading is done by pulling a wedge to spread a sample drop of blood on the slide.
A well-made peripheral smear is thick at the frosted end and becomes progressively thinner toward the opposite end.
In the thicker region, most of the cells are clumped, which increases the difficulty in identifying and analyzing blood components.  At the thinner region the cells are unevenly distributed, and grainy streaks, troughs, ridges, waves, or holes may be present. This portion
of the smear has insufficient useful information for analysis. The "zone of morphology" (area of optimal thickness for light microscopic examination) occupies the central area of the slide. Figure \ref{sangre1} shows typical images captured on the same peripheral blood smear.

The thickness of the smear is influenced by the angle of the spreader (the wedge), the size of the drop of blood, and
the speed of spreading.  In current laboratory practice, skilled users manually identify
the \textit{zone of morphology} and acquire images of it. Due to the aforementioned
reasons, this zone varies in its morphology and offers specific appearances
on different slides.  Such manual
identification is tedious, inconsistent, and prone to error, and is
also biased in terms of statistics and user subjectivity.  Advances in high-throughput microscopy have enabled the
rapid acquisition of many images without
human intervention. Depending on the sample
size on the slide, one could easily acquire more than ten
thousand images in a sample. An automatic detection of the "zone of morphology" can increase consistency,
reduce labor, and achieve better accuracy.
Some papers in the literature about automatic
classification of this area are  \citep{Mutschleretal87,anguloetal03}.

Our dataset consists of three peripheral blood smears that were obtained from
patients with Sickle Cell Disease.   Thirty digital images were taken across each smear (some of them are shown in Fig. \ref{sangre1}). Our aim is to use the inhomogeneous $K$-function to classify these images in three groups, which would correspond to the thick, thin and morphological zones respectively.

\begin{figure}
   \begin{center}
\begin{tabular}{ccc}
\includegraphics [width=3.5cm]{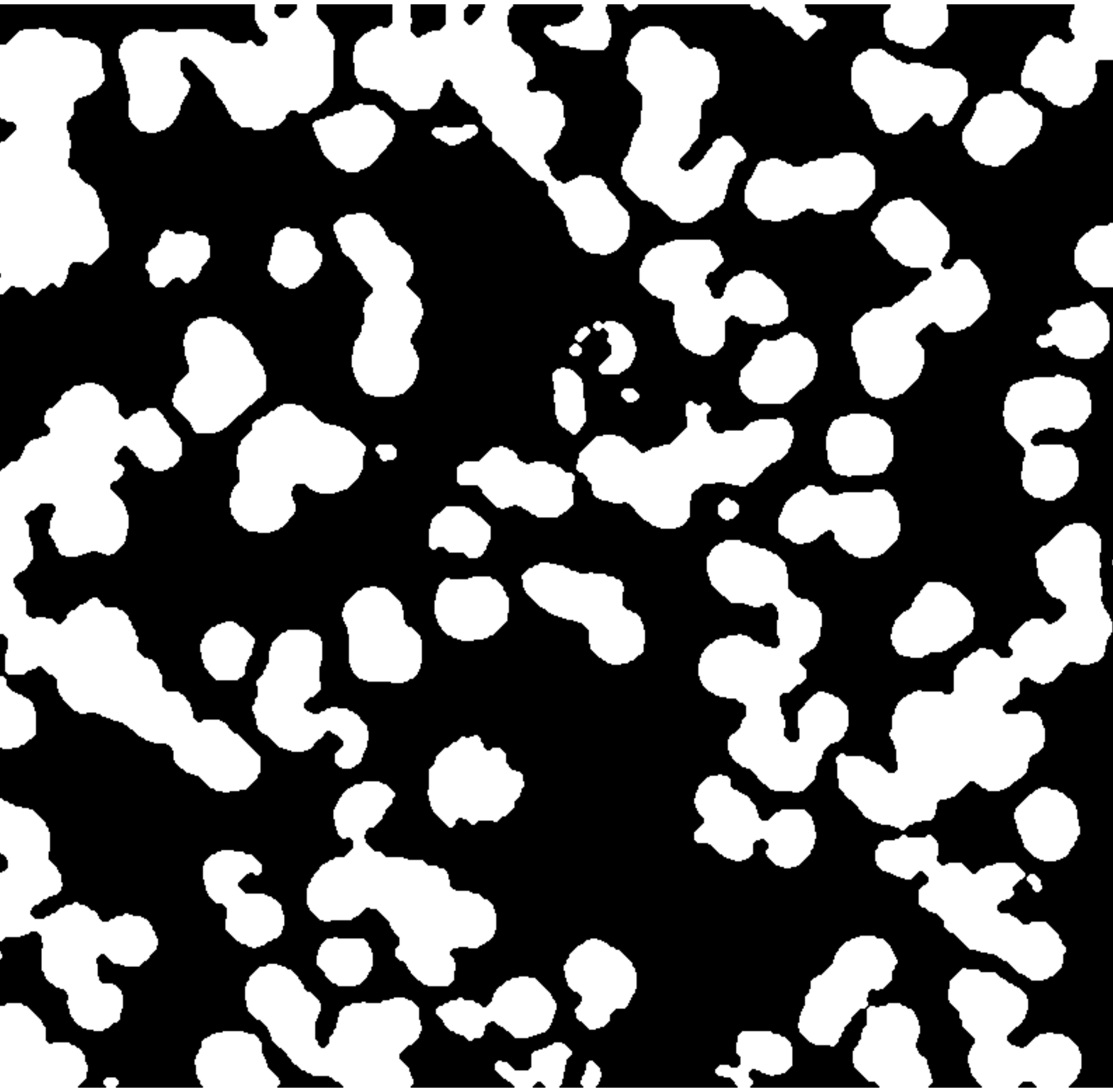} &
\includegraphics [width=3.5cm]{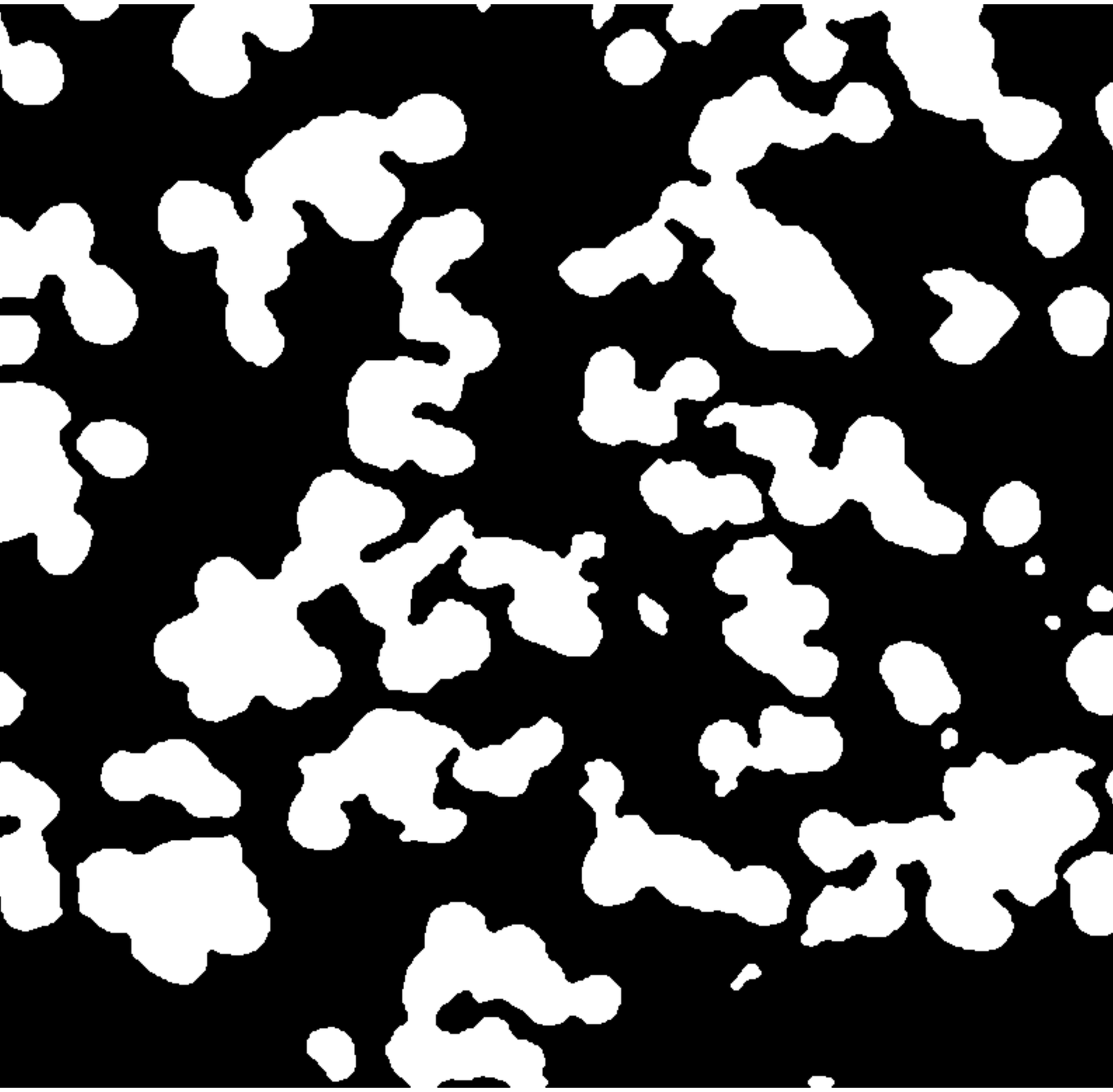} &
\includegraphics [width=3.5cm]{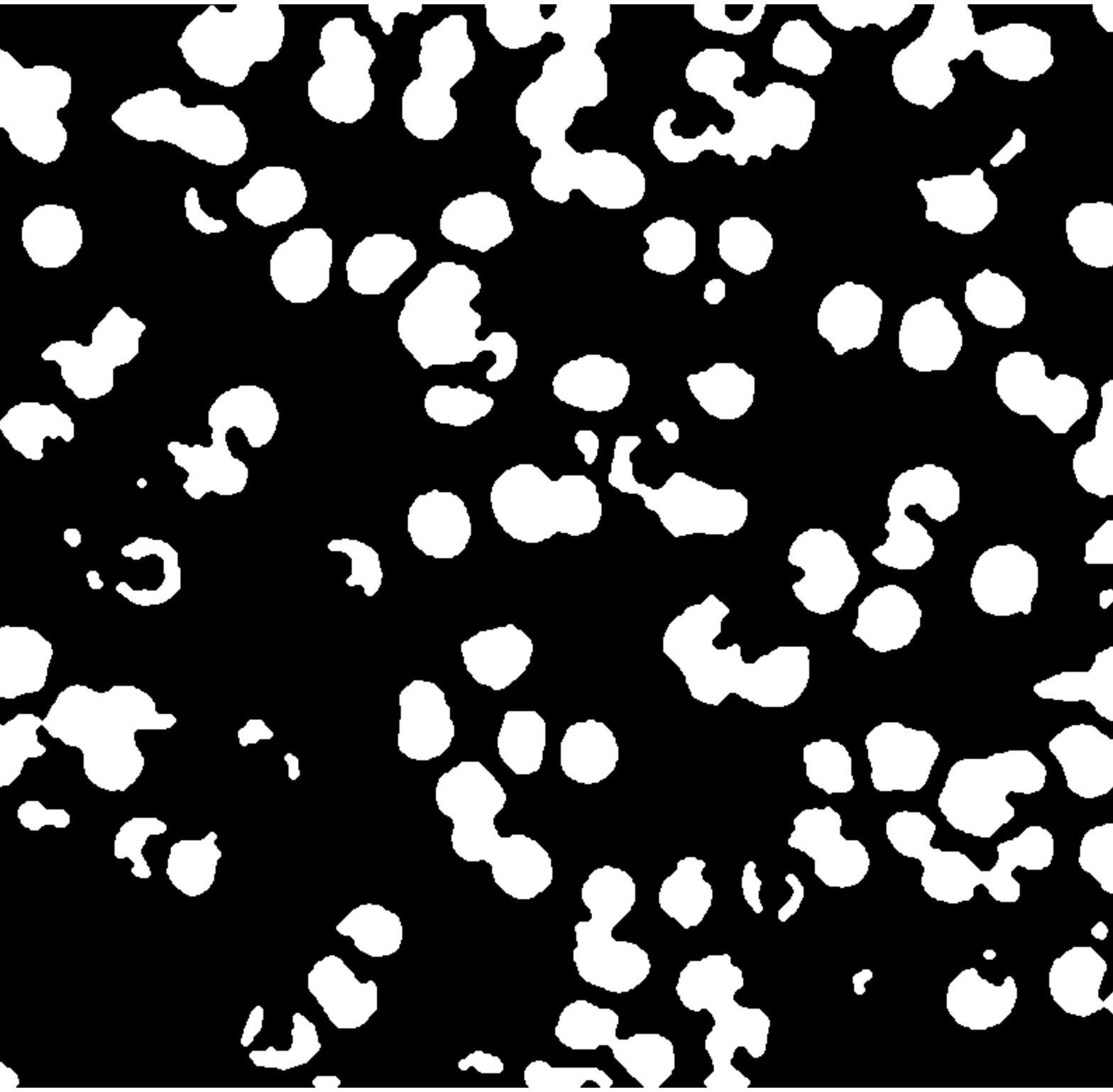} \\
\includegraphics [width=3.5cm]{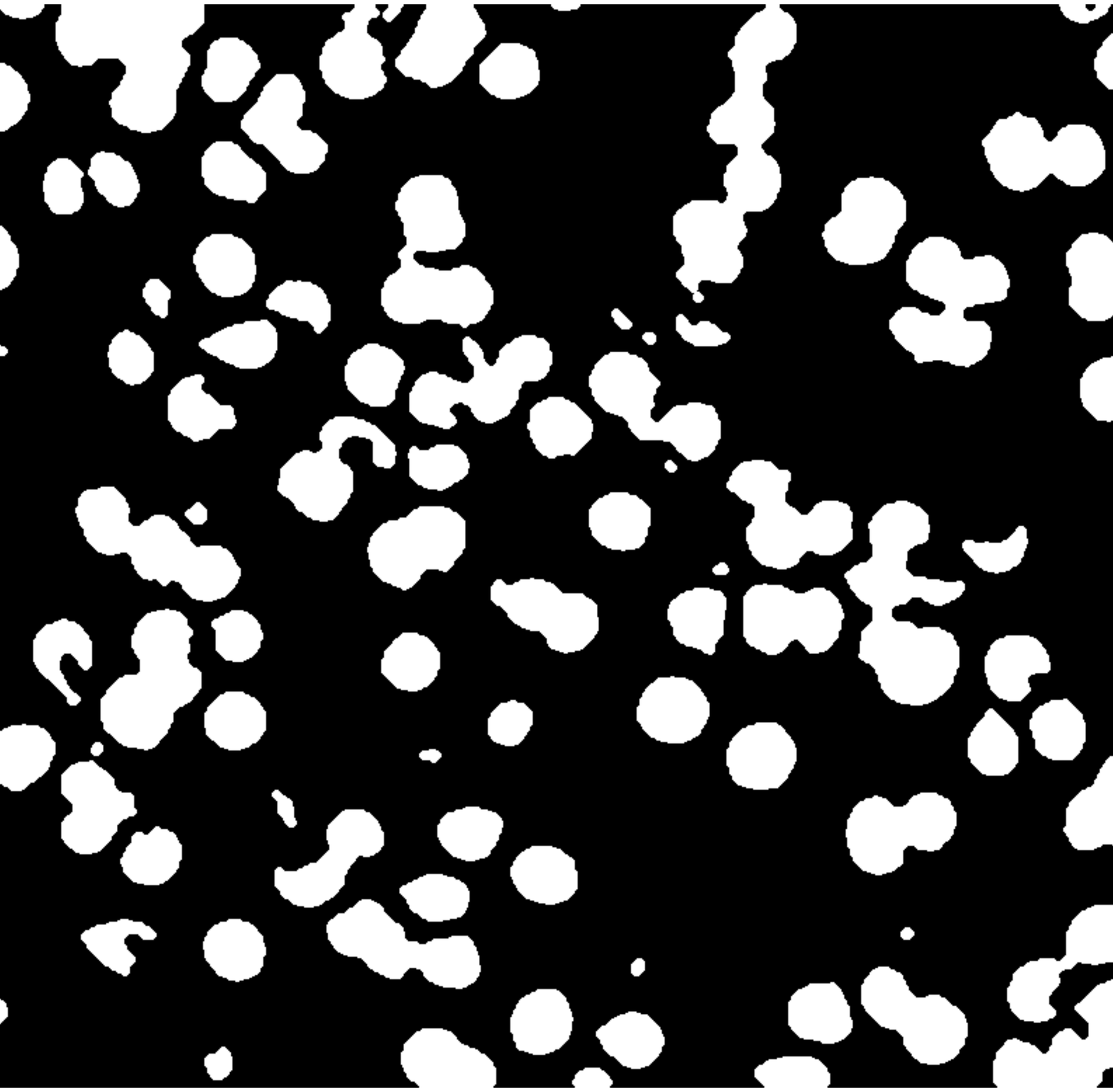} &
\includegraphics [width=3.5cm]{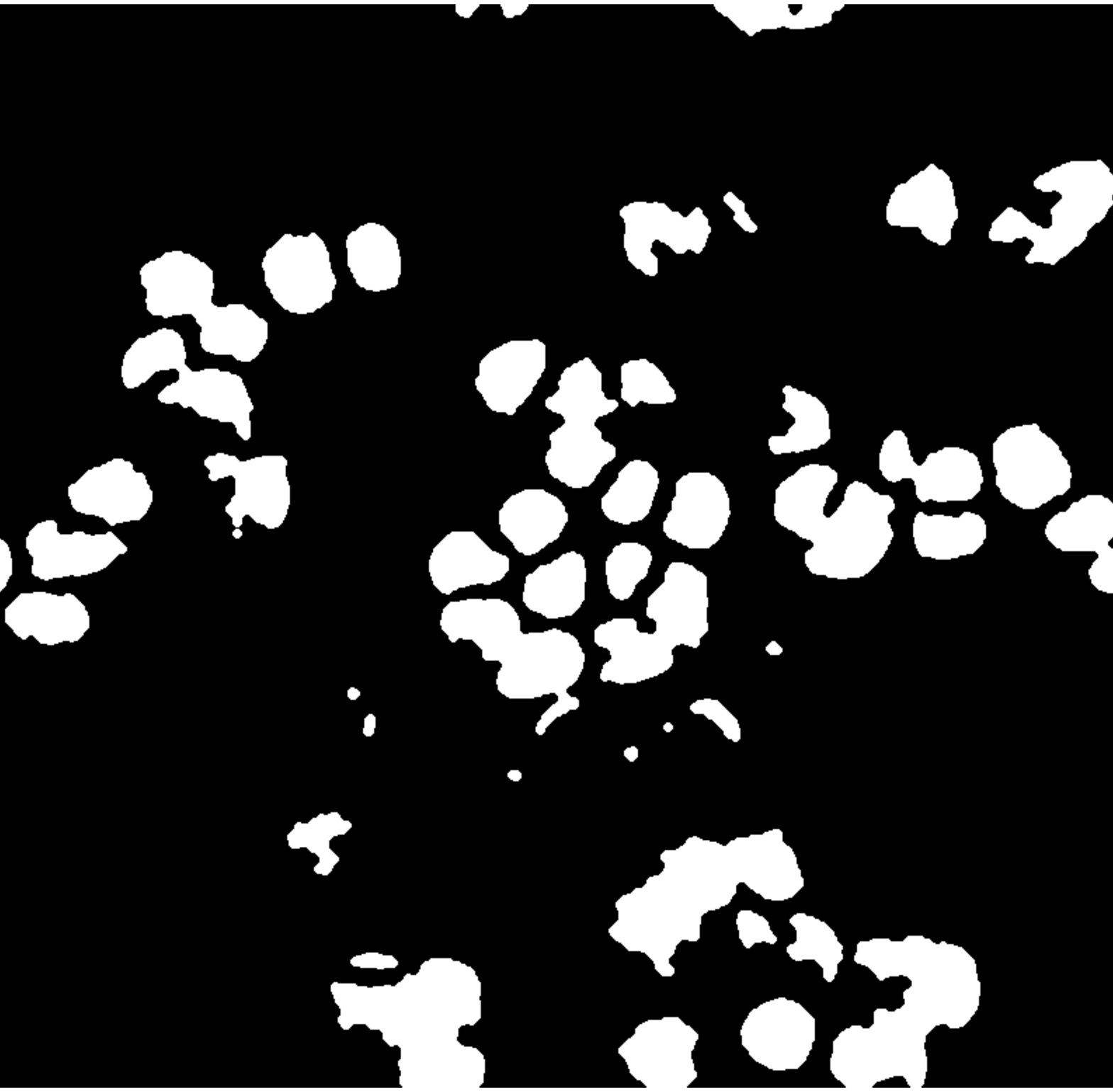} &
\includegraphics [width=3.5cm]{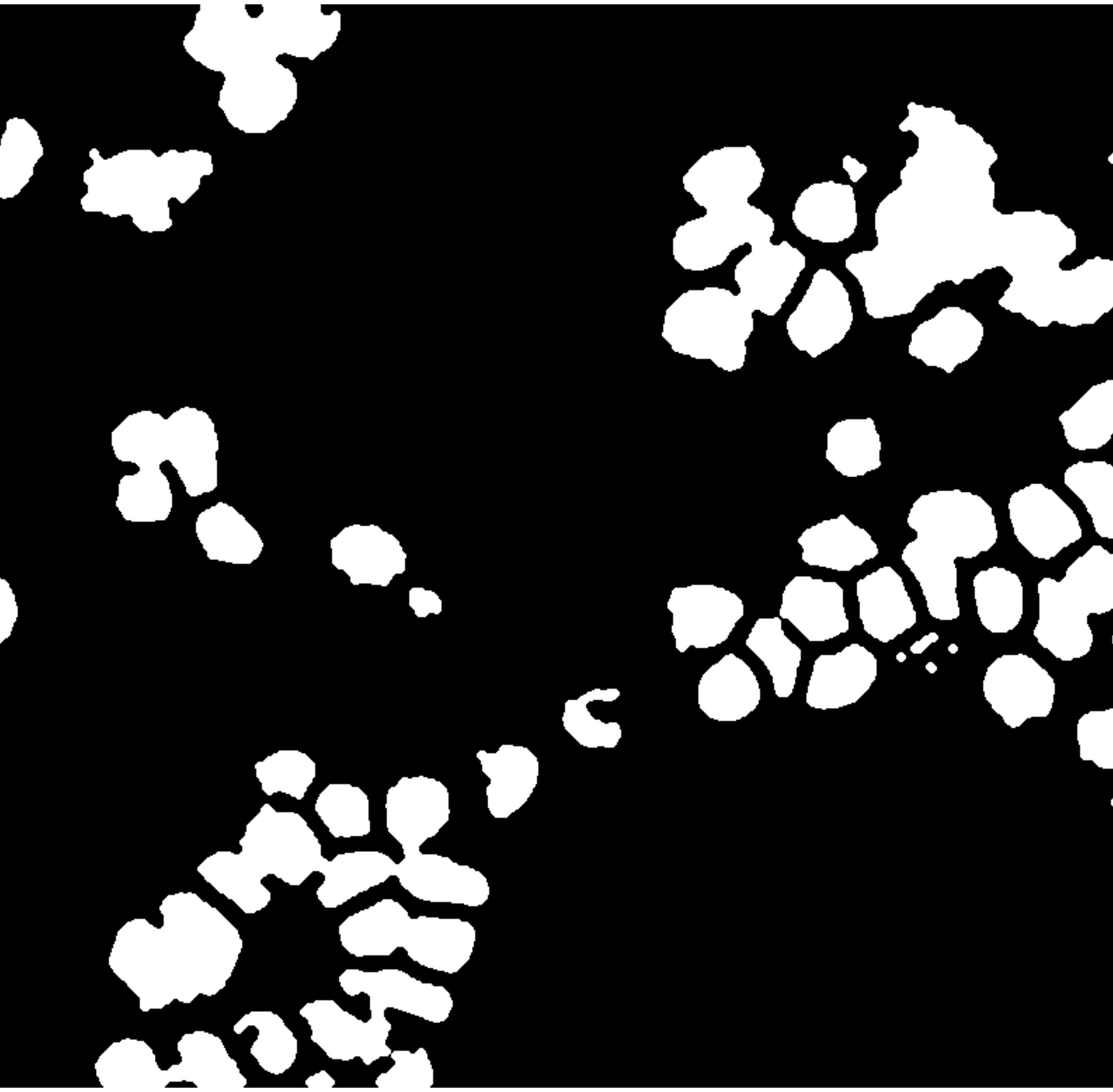} \\
 \end{tabular}
  \end{center}
  \caption{ Pre-processed images of different zones of the same peripheral blood smear \label{sangre2}}
\end{figure}

Prior to proceeding with the clustering, the original images are segmented in order to convert them into binary images.  Since there exists a good contrast between cells and background, a good segmentation is obtained using very simple image processing techniques:  thresholding followed by morphological filtering. This binarization was performed using Matlab. In Figure \ref{sangre2} we can see the binary images corresponding to the images shown in Figure \ref{sangre1}.
These binary images are considered realizations of three different germ-grain processes, so the inhomogeneous K-function can be used to define homogeneous classes of images. In order to do that, the inhomogeneous K-function is estimated from each image (Eq. \ref{Kin3}), and a partitioning method called PAM (Partition Around Medoids) \citep{Kaufman90} will be used for clustering, defining the distance between each couple of images  as the Euclidean distance between their inhomogeneous K-functions.

\begin{figure}
   \begin{center}
\begin{tabular}{ccc}
\includegraphics [width=3.7cm]{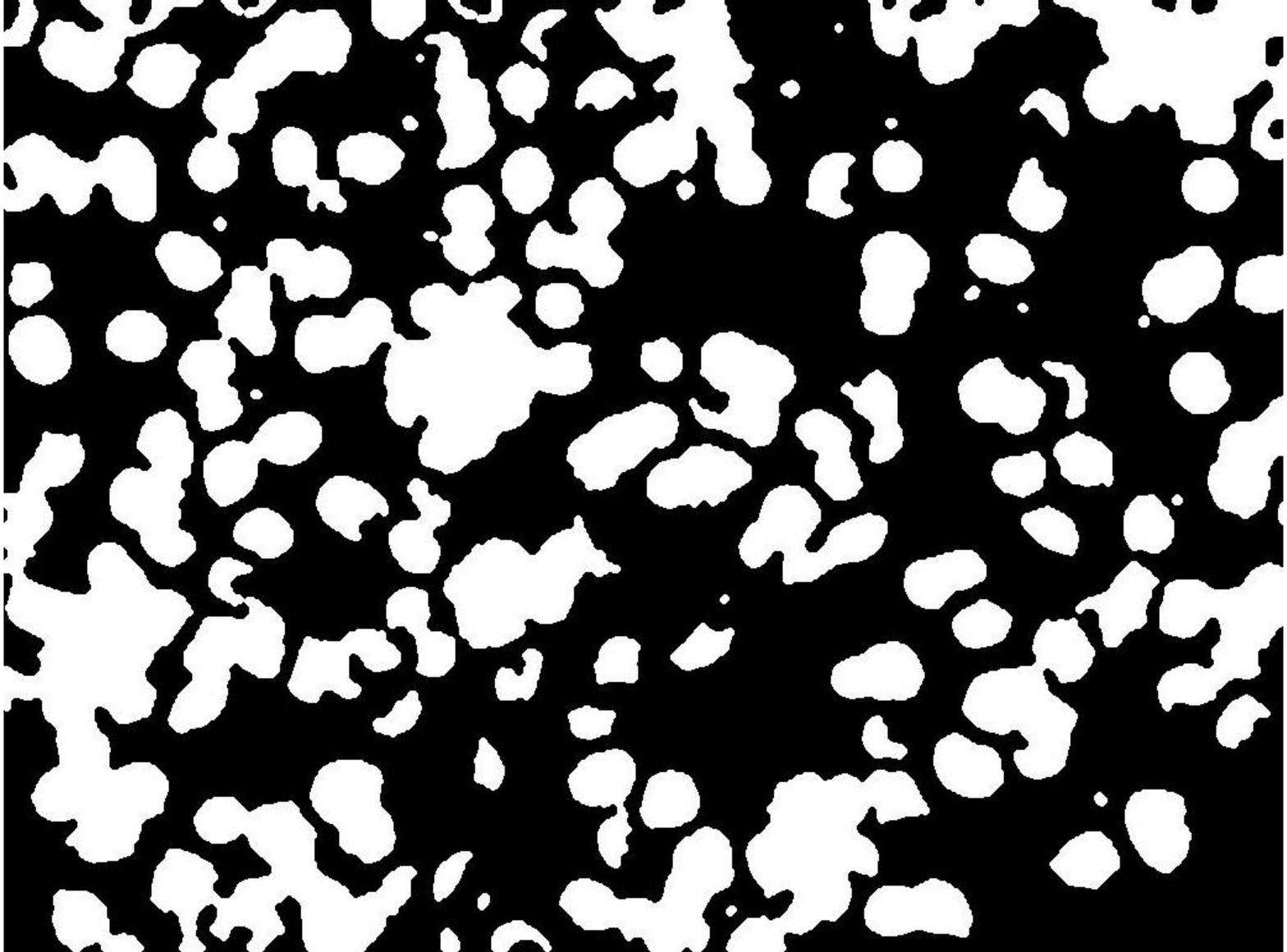} &
\includegraphics [width=3.7cm]{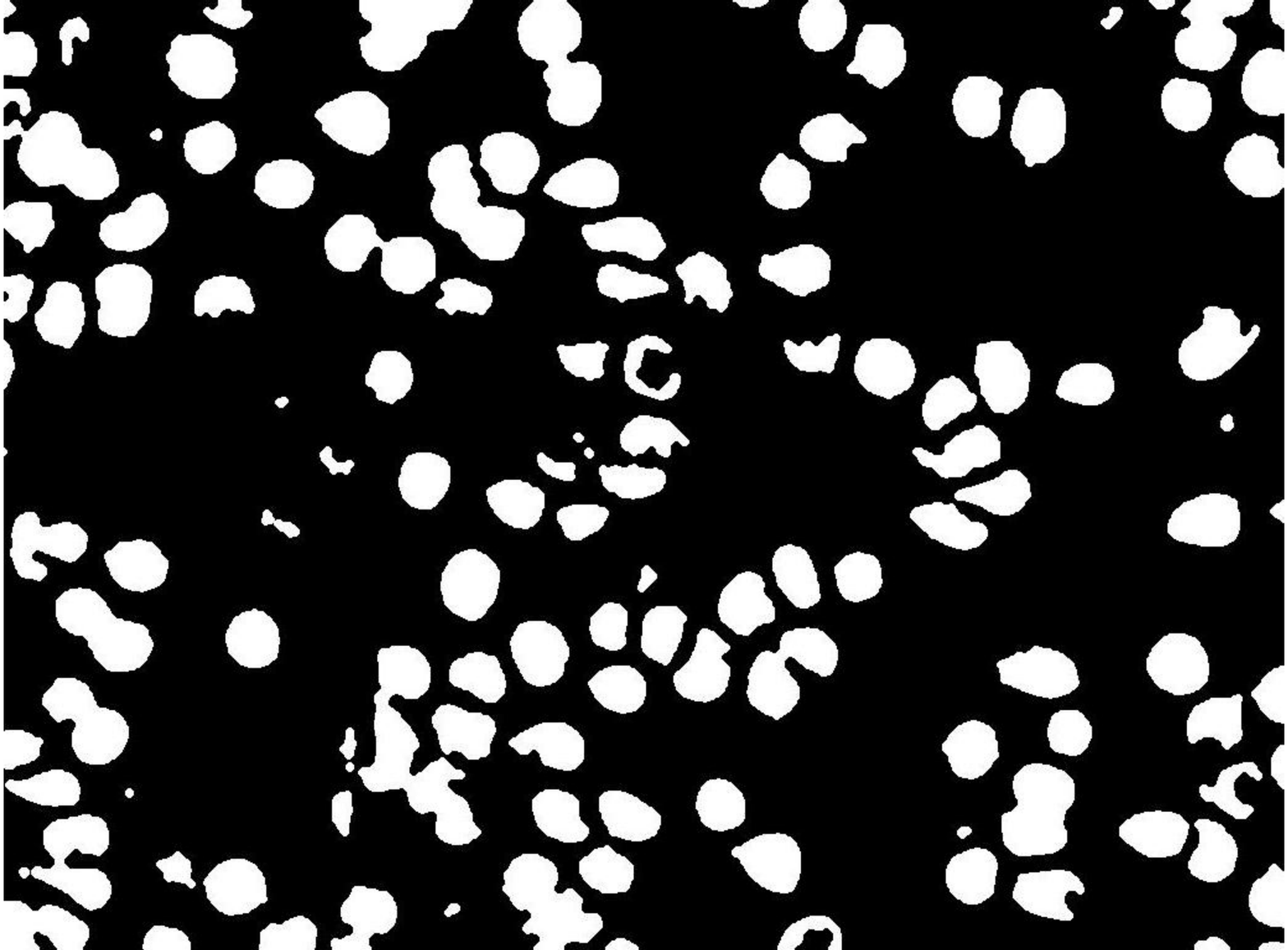} &
\includegraphics [width=3.7cm]{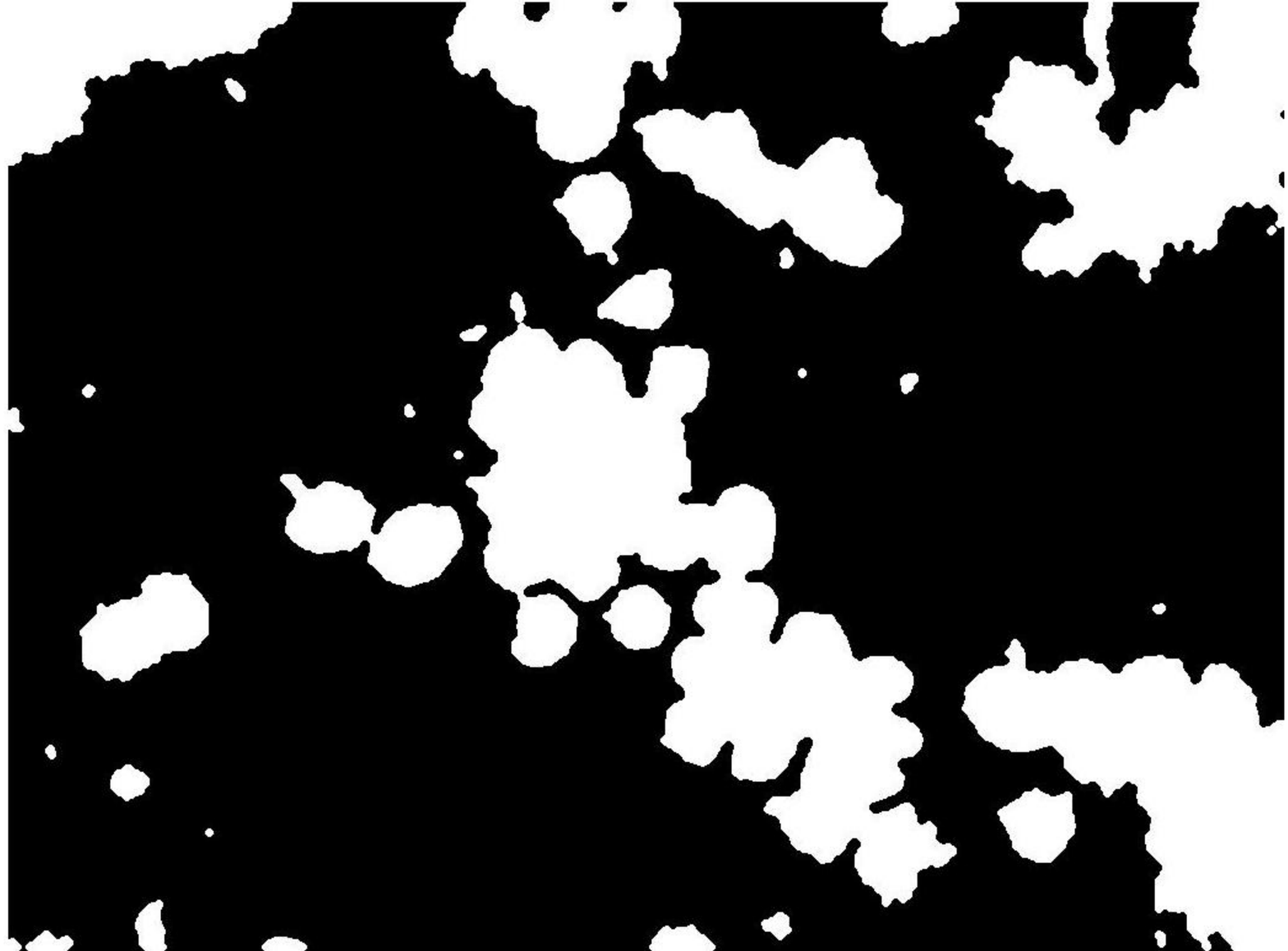} \\
 \end{tabular}
  \end{center}
  \caption{ Medoids of the three clusters obtained using the Euclidean distance between the K-functions \label{medoidesK}}
\end{figure}

Fig. \ref{medoidesK}  shows the images corresponding to the medoids obtained for the three clusters.

After obtaining the clusters, an expert hematologist helped us to review the results and confirm the goodness of the classification obtained by looking at the medoids and the other images of each group.
It is very important to remark that this study should be carried out with a really large
image database in order to obtain any valid clinical conclusions, but in any case our results are very
promising.

\section{Conclusions} \label{conclusiones}

In this paper we
 introduce a generalization of the inhomogeneous K-function that allows its application to non-homogeneous germ-grain models. We have shown its capacity to discriminate between realizations of different models and we have applied it to a clinical application: a sample of images of
peripheral blood smears obtained from patients with SCD.  Different groups of images corresponding with different patterns were found. The study should be repeated with a really large image database in order to achieve valid medical conclusions. From our
point of view, we believe our methodology can be used without modifications.
Other future applications of the inhomogeneous K-function can include Monte Carlo goodness of fit test \citep{Diggle83} or parameter estimation.


\section{Acknowledgements}
We would like to thank Silena Herold from  the Computation Faculty of the Universidad de Oriente, Santiago de Cuba, for introducing us in this interesting problem and providing us the images.

This work  has been partially supported by  the UJI projects  $P11B2012-24$  and $P11A2011-11$.

\bibliographystyle{plainnat}

\end{document}